\documentstyle[prl,aps,preprint]{revtex}
\begin{document}
\tightenlines
\preprint{GTP-98-6}
\title{Correspondence between classical dynamics and
  energy level spacing distribution in the transition
  billiard systems}

\author{Soo-Young Lee and  Sunghwan Rim}
\address{Basic Science Research Institute, Korea University,
Seoul 136-701, Korea}

\author{Eui-Soon Yim}
\address{Department of Physics, Semyung University, Chechon
390-230, Korea}

\author{C. H. Lee}
\address{ D\&S Dept.,R\&D Centernam Semiconductor Inc.,
Seoul 133-120, Korea}

\date{\today}
\maketitle
\begin{abstract}
The Robnik billiard is investigated in detail both classically
and quantally in the transition  range  from integrable
to almost chaotic system. 
We find out that a remarkable correspondence between 
characteristic features of classical dynamics, especially
topological structure of integrable regions in the Poincar\'{e}
 surface 
of section, and the statistics of energy level spacings
appears with a system parameter $\lambda$ being varied.
It is shown that the variance of the level spacing  
distribution changes its behavior at every particular values 
of $\lambda$ in such a way that
classical dynamics changes its topological structure in
the Poincar\'{e} surface of section, while the skewness and 
the excess of the level
spacings seem to be closely relevant to the interface structure
between integrable region and chaotic sea rather than
inner structure of intergrable regoin.
\end{abstract}


\vspace{2cm}

It is very important to know what characteristic properties 
in quantum mechanics represent classical chaos.
Many authors have devoted their efforts on this subject\cite{Gu90,Hu97}.
It is, now, well known that for K-system its level spacing
distribution has a universal form, i.e., Wigner
 distribution\cite{Za81}.
However, for a soft chaotic system, any universal property
has not been reported so far. In semiclassical limit
Berry and Robnik\cite{Be84} suggested a level spacing distribution
 which contains a physical 
parameter $\rho_{cl}$, the phase space volume of integrable
 region
in the soft chaotic system. This semiclassical distribution
is, however, not confirmed completely by numerical or
 experimental
 analyses\cite{Ho89,Pr93}.
Recently, Rim {\it et. al.}\cite{Ri98} show numerically
 in the Dreitlein billiard that the level spacing
distribution
is very sensitive to a  small change of the system parameter
in the  vicinity of changing points of classical phase space
topology like  bifurcation points of robust islands.
 This result is very important
because it  implies that a universal distribution for
a soft chaotic system,
 if it would exist, should contain a physical parameter
related to the topology of the classical phase space manifolds
 in addition to $\rho_{cl}$.

In this Letter, we carefully investigate the Robnik billiard 
classically and quantally in the soft chaotic region of
 $\lambda$,
and confirm the sensitivity of the level spacing statistics
to the topological change of phase space  manifolds 
 for the corresponding classical system.
Furthermore, we try to find out a detailed correspondence
between behaviors of variance, skewness, and excess of
the level spacing distribution and changes of classical
phace space structures.
The Robnik billiard has been
studied by many authors in soft chaotic range as
 well as
hard chaotic range\cite{Li96,Li98,Li94,St93}.
 However, it has not been
 analysed in this view point.

 The Robnik billiard is given
as a quadratic conformal map $w=A z + B z^2$ of the unit
circle, and the area $S=\pi ( A^2 + 2 B^2)$ is fixed as $\pi$.
The system parameter $\lambda$ is related to $A$ and $B$ as
\begin{equation}
A = \cos p, \,\,\, B=(1/\sqrt{2}) \sin p, \,\,\,
p=\tan^{-1} (\lambda \sqrt{2}).
\end{equation} 
The Robnik billiard has several adventages for our purpose.
First, it is continuously transformed with increasing $\lambda$
 from an integrable circle system to an almost chaotic system
through soft chaotic systems which have both integrable 
parts and chaotic parts. This property enables us to 
investigate
the soft chaotic system varying continuously with $\lambda$.  
Second, it has an analytic boundary, while the Dreitlein 
billiard
has non-analyticity in the boundary when the system is  soft
 chaotic,
 so that the integrable parts of the Poincar\'{e} section have 
complex
 structures which change continuously with $\lambda$.
This property may give a partial support for generality of
the sensitivity of the level spacing statistics. 
Third, the method of calculating  energy levels is 
already given so that the energy level calculation
is easily performed by the diagonalization method\cite{Ro84} which
is known to be very accurate.
 Final advantage is that the system with larger than $\lambda=0.2$
has only one dominant island surrounding period two orbit in 
chaotic sea. This means that without effects of other islands
we can analyse impacts of the structure change of the island
on the statistics of level spacings.

Classical analysis on the Robnik billiard has been performed and
several bifurcation points are reported\cite{Ro83,Ha87}.
 We also investigate 
the Poincar\'{e} surface of section and confirm
those points. Additionally we find a new pattern of the structure 
change of island within soft chaotic range of $\lambda$.
We summarize results of the analyses for the classical dynamics
 below.
 At about $\lambda = 0.175$,  three kinds of  islands dominate  
the Poincar\'{e} section, which are corresponding to period two,
 period three, and
period four orbits. The period four orbit is bifurcated 
at $\lambda = 0.176$ and the period five
 orbit  at $\lambda =0.185$.  
After then, the only island surrouding
period two orbit remains alone in chaotic sea, and the period
two orbit is bifurcated at $\lambda = 0.207$. 
As usual, this bifurcation does not mean that the island is
divided into two isolated islands by chaotic sea.
Just after the bifurcation the bifurcated orbits are still
 wrapped
by invariant tori as described in Fig. 1 (a).
These invariant tori break out  at about $\lambda = 0.219$ 
so that
two isolated islands centered at the bifurcated orbits appear
in chaotic sea(Fig. 1 (b)).
These bifurcated
orbits are again bifurcated at $\lambda =0.266$ which is
 reported by Hayli {\it et. al.}\cite{Ha87}. 

In order to see the structural change of the island   
surrouding period two orbit, we plot carefully
 the island in the range of $ 0.19 < \lambda < 0.25$ where
the structures of integrable parts are effectively given  by 
those of the period two island.  
From this plotting we find that the structure of the period
two island evolves after a certain pattern
with increasing $\lambda$. The pattern is shown in Fig. 2.
The first step of the pattern is that resonances appear in 
mid of the 
island, and secondly, the size of resonances grow gradually 
and the position  goes to 
outer part of the island.  As the final step,  all invariant
 tori wrapping the 
resonances break out so that the resonances become independent
 islands embedded in chaotic sea and, then, these new islands
 disappear rapidly.
This pattern is repeated several times in the range of 
$ 0.19 < \lambda < 0.25$.
We note that the final step of the pattern is a topological
changing point
of the interface structure between the island and chaotic sea.
 This point as well as bifurcation points would play
an important role in understanding the correspondence 
between classical chaos  and quantum chaos.
In practice, it is not simple to determine numerically 
the precise value of $\lambda$ at which such break of
invariant tori appears. We, therefore, assume that
not all invariant tori wrapping the resonances would
be broken if a trajectory starting from the thin chaotic
region between the resonances and the island does not
reach chaotic sea within 10000 boundary collisions.
  Using this way, we can determine
such  breaking points as $\lambda = 0.203,0.227,0.231, 0.238$.
In addition to the pattern, we find another
structure change of the island at $\lambda =0.245.$
Before and after this point the shape of island is
inverted like mirror images, and right at the point
the size of island becomes very small as shown in Fig. 3.
As explained above the Robnik billiard has various
structural behaviors compared with the Dreitlein billiard
where the period two island does not show such complex
 pattern  due to the simplicity 
 and non-analyticity of the boundary.
 This relatively complicated behavior of
 robust island structure enables us to investigate a
 detailed correspondence between classical phase space
structure and quantal level spacing statistics.

Using the diagonalization method\cite{Ro84},
we  obtain energy eigenvalues for the Robnik
 billiard.
We calculate 1100 energy levels at discrete values of
 $\lambda$
with an interval $\Delta \lambda =0.001$.
 Among  1100
energy levels the lowest 500 levels, which are assured
 to be reliable,
 are taken for calculating the level
spacing statistics. Of cource, calculation of further 
energy levels is desirable and it may give 
more precise statistics.
We, however, believe that the lowest 500 levels are 
enough to see 
characteristic behavior of level spacing statistics
along $\lambda$. 

We obtain the Brody distribution exponent $\nu$ making
a comparison with the normalised variance $\sigma^2$ 
of the level spacings as
\begin{equation}
\sigma^2 =2 (\nu+1)\Gamma (2/(\nu +1))
/[\Gamma(1/(\nu +1)) ]^2 -1.
\end{equation}
This method was used by Robnik[ref].
The results are shown in Fig. 4.
The skewness and the excess are also obtained and shown
 in Fig.5.
 In Figs. 4 and 5, the vertical dotted lines indicate
the $\lambda$  values at which the wrapping invariant
tori break out, and the dashed line does the structural
changing point of island at $\lambda = 0.245$.
We omitted the line of $\lambda=0.231$ because this point
seems to give a minor effect on the statistics.
The vertical arrows point out the bifurcation points
of period four, three, and two orbits.
As shown in Fig.5, the exponent $\nu$ increases
globally up to about $\nu =1$ indicating
the transtion from the Poisson to the Wigner
distribution.
However, it can be seen clearly
that the detailed glimpse of the increasing
behavior discloses a piece of veil;
it is not gradual and smooth, but is rather 
 staircaselike.
This behavior implies that
there must be some cmpeting process independent
of the grobal parameter $\rho_{cl}$ which
alone would smoothly increases 
 the variance of level spacings with increasing 
$\lambda$.
It is surprising that
the starting  points showing steep increase of $\nu$
are located on the points indicated by the dotted lines
 or arrows. Since the dotted lines and arrows denote
the positions
at which structual changes of islands appear in
classical phase space, these coincidence strongly
suggest that the level spacing statistics
are affected by the topology of the phase space
manifolds as well as the grobal parameter $\rho_{cl}$
for the corresponding classical system.  
This staircaselike behavior is very similar
 to the case of first order
phase transition; temperature($\nu$) does not change
even under external injection of energy($\lambda$)
while the phase of matter(topological structure of 
classical phase space) is changing.
More plausible physical situation for this seems that
changing procedure of the phase space structure, i.e.,
occurrence and growing of resonances, may give
the effect of lowering the exponent $\nu$,
and whose effect would end at topological
changing point of the phase space manifolds. 

The plots for the skewness and the excess
 of the level spacing
distribution also give a similiar correspondence with
classical dynamics.  Similar behaviors with $\lambda$
are shown in the skewness plot and the excess 
plot(Fig. 5).
These plots have several local maxima and show
tendency to globally decrease up to about 
$\lambda =0.25$. It is worthy to note the dotted lines,
which indicate the topological change of interfaces
of island with chaotic sea,
are located at local mimima while the bifurcation
point of the period two orbit is not coincident
with any maximum or minimum positions.
It gives an evidence for the fact that the skewness
 and
the excess are sensitive to the outer structural
change of island rather than inner change such as
 bifurcation point. 
Another interesting feature of the level statistics
is that, except the plateau just before the bifurcation
point, all plateaus of $\nu$(or variance) are located
on the rapid  decreasing range after maxima
in the skewness and the excess.
From this observation we may speculate that during
the procedure of topological change of island 
the $\nu$ (or variance) has the almost same value
and the skew and the excess are decreased rapidly.

In the conclusion, 
we show the correspondence between classical 
dynamics, particularly, the topology of the
phase space manifolds, and energy level spacing
statistics for the soft chaotic range in the
Robnik billiard.
The topological change of phase space manifolds 
delivers a sensitive impact upon the level spacing 
distribution. The variance of the distribution
seems to be affected by every topological change,
while the sknewness and the excess have higher
sensitivity to the outer structural changes
 of integrable parts than inner changes.

It is natural and important to raise the question
as `Does this correspondence appear
in the statistics for very high energy levels?'.
A clear answer for this question would be given
by future work. Our conjecture tells that it would
be  the case even for semiclassical limit.

The authors would like to express many thanks
to  other GTP members, C.S. Park, D.H. Yoon, S.K. Yoo,
 and D.K. Park, for very sincere and informative 
discussions.

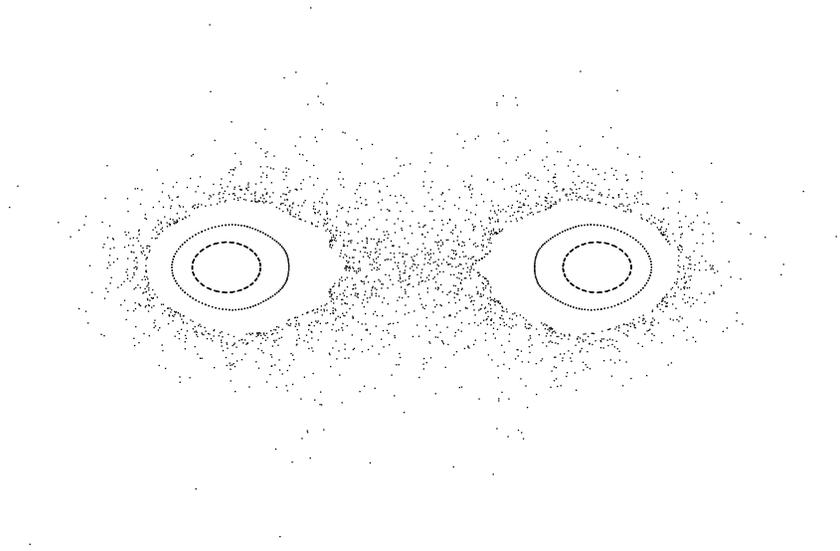
\begin{figure}

\caption{ (a) $\lambda = 0.210$. \,\, (b) $\lambda = 0.220$. 
}
\end{figure}

\begin{figure}

\caption{(a) $\lambda = 0.210$. \,\, (b) $\lambda = 0.202$.
\,\, (c) $\lambda = 0.205$.
}
\end{figure}

\begin{figure}

\caption{(a) $\lambda = 0.242$. \,\, (b) $\lambda = 0.245$.
\,\, (c) $\lambda = 0.248$.
}
\end{figure}

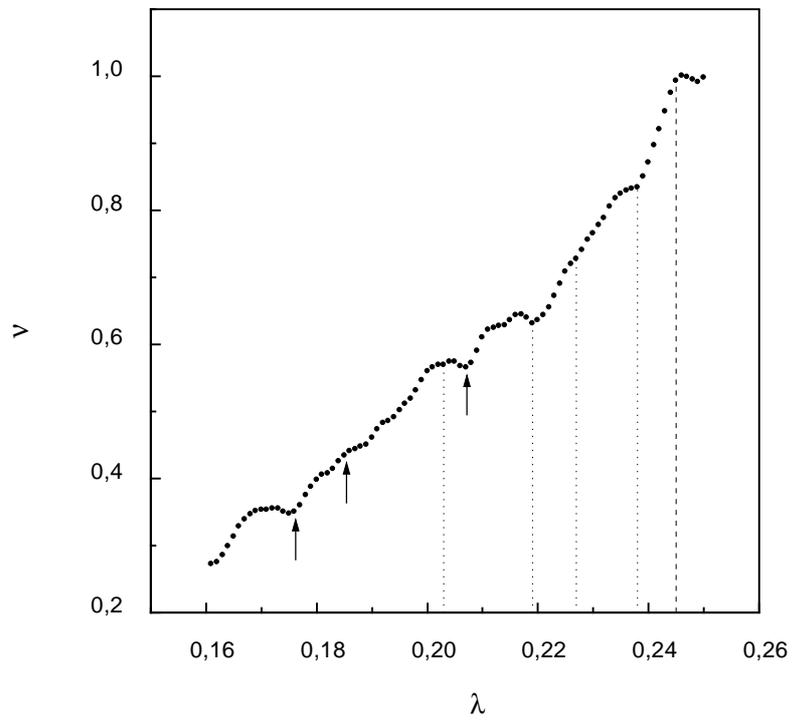
\begin{figure}

\caption{ The Brody exponent $\nu$ versus $\lambda$. 
}
\end{figure}

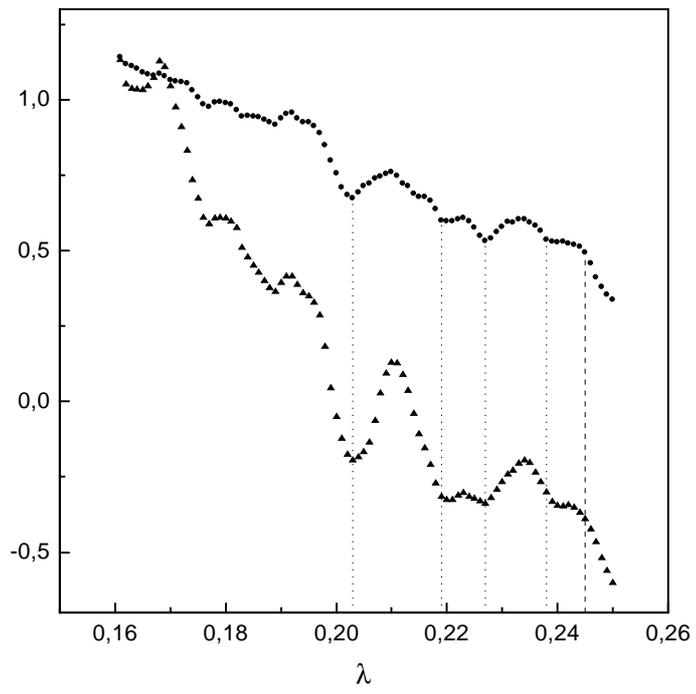
\begin{figure}

\caption{ The skewness(circles) and the excess(triangles) 
with $\lambda$.
}
\end{figure}

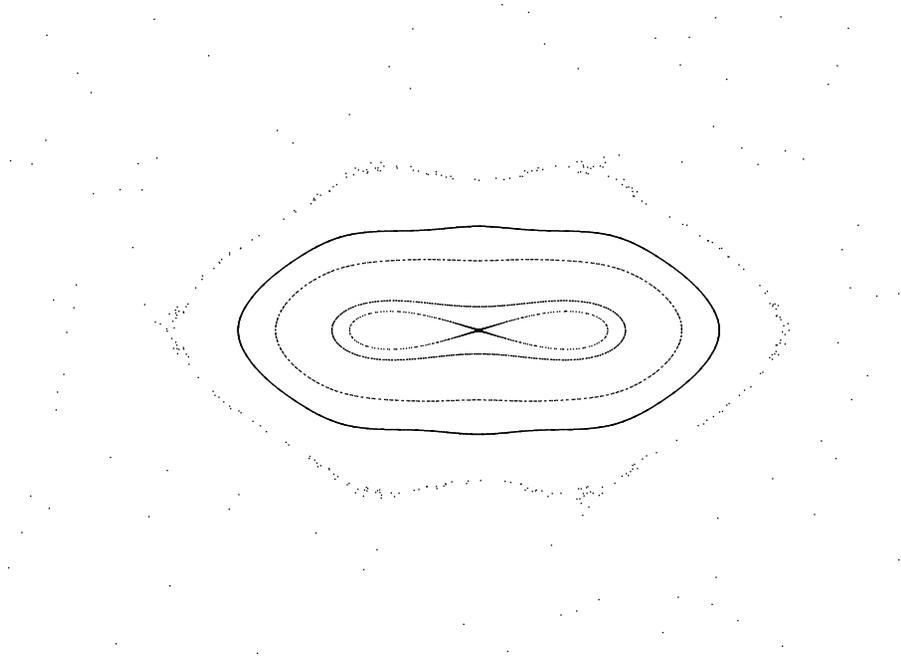
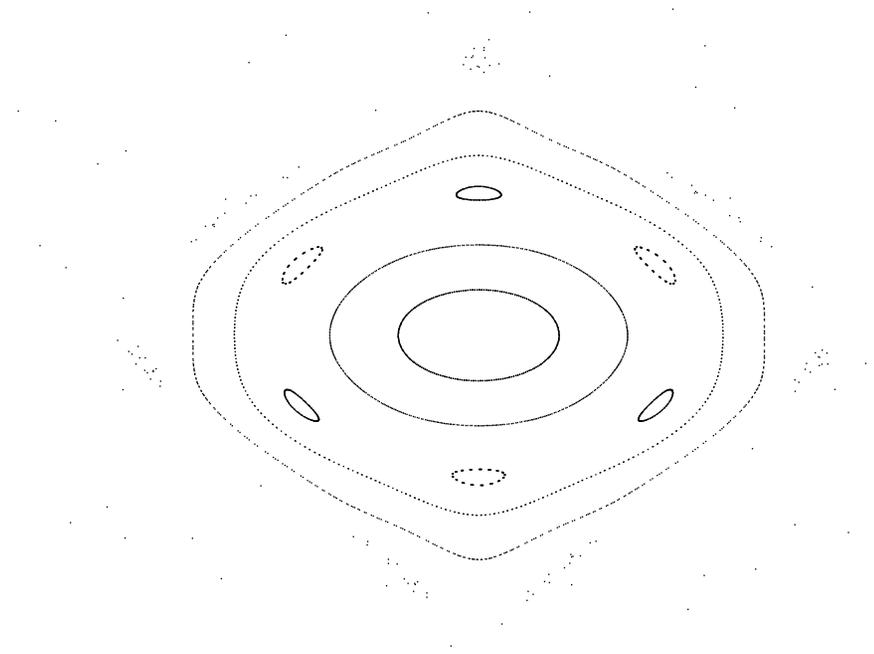
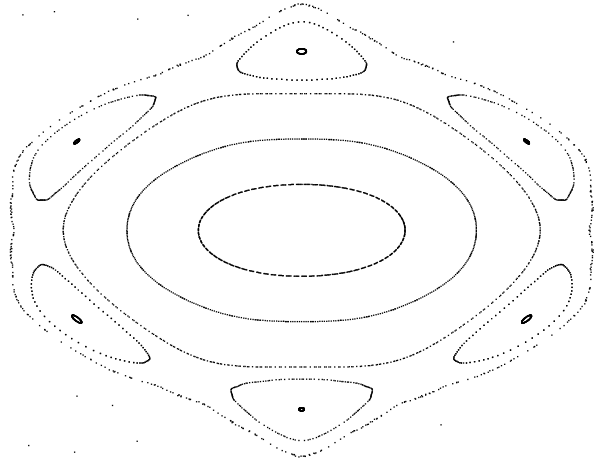
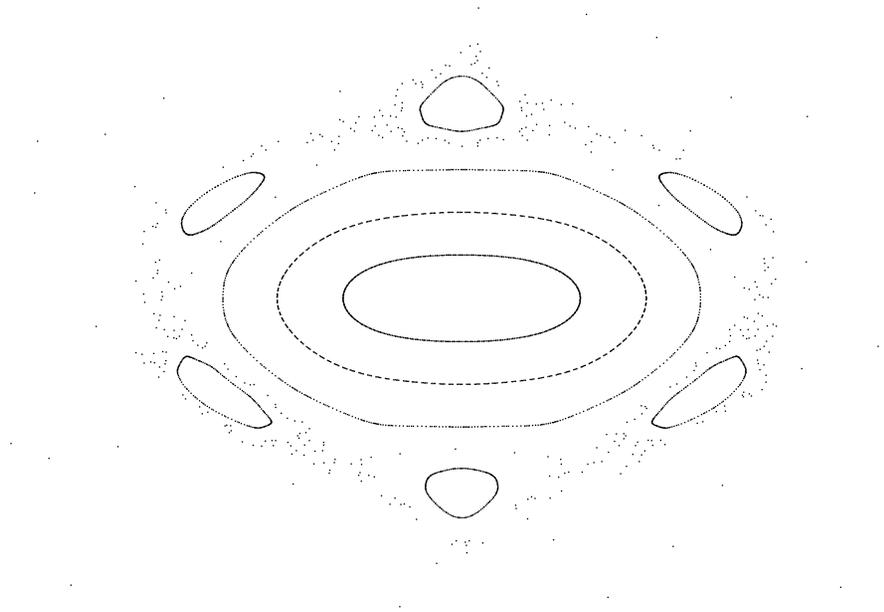
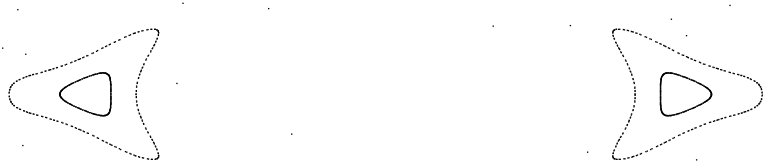
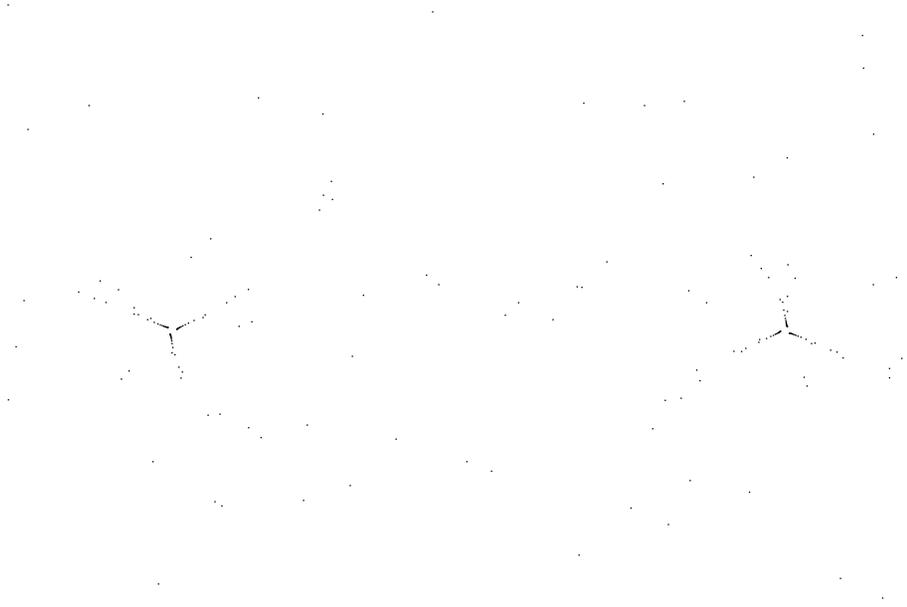
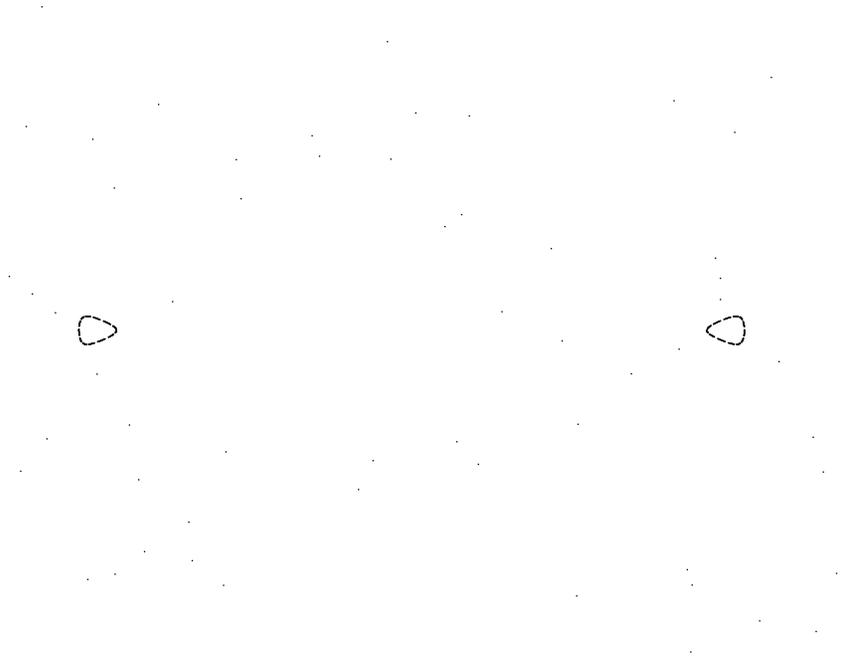
\end{document}